# SISTEM INFORMASI PENJUALAN DAN PERBAIKAN KOMPUTER
(Studi Kasus: CV Computer Plus Palembang)

Oleh: Syaprina, Leon Andretti Abdillah, & Nyimas Sopiah
Mahasiswa & Dosen Universitas Bina Darma, Palembang

***Abstracts***: *The purpose of this research is to develop an Information System of Selling and Services using Microsoft Visual Basic and Microsoft Access for it database. The benefits of this research is to help CV Computer Plus in selling and services data processing everyday. To develop this IS is used 5 (five) steps: 1) Planning, 2) Analysis, 3) Design, 4) Implementation, and 5) Evaluation. The Information System can record the selling and services data, it also prepared usefull reports. By using this IS, CV Computer Plus can operate their selling and services efficiency and effectively. In the future it can be upgraded for network application.*

***Keywords****: Information System, Selling, Services.*

## 1. PENDAHULUAN

Perkembangan teknologi informasi dari tahun ke tahun yang semakin cepat menjadi tantangan berat bagi pengguna teknologi informasi itu sendiri, dan mendorong setiap sektor organisasi baik formal maupun informal atau lembaga-lembaga lainnya untuk dapat memanfaatkannya sebagai penunjang kegiatan kerja sehingga dapat menghasilkan informasi yang cepat, tepat dan akurat. Untuk mewujudkan hal tersebut, maka dibutuhkan sumber daya pendukung lainnya seperti perangkat lunak yang dapat diandalkan kemampuannya serta sumber daya manusia yang harus menguasai kemampuan teknologi informasi itu sendiri. Dari perkembangan teknologi itulah kita harus memahami serta mengenal teknologi tersebut. Yang mana kecanggihan teknologi akan terus berkembang dengan pesat diberbagai aspek kehidupan di masa yang akan datang.

CV Computer Plus Palembang merupakan salah satu perusahaan swasta yang bergerak di bidang penjualan dan perbaikan komputer. Selain menjual komputer CV Computer Plus juga menjual *accesories* komputer seperti *flashdisk, mouse, keyboard, catridge*, *printer*, *ink* dan lain-lain.

Pada bagian penjualan di CV Computer Plus dalam melakukan pengolahan data penjualan sehari-hari saat ini penggunaan komputernya belum optimal. Dimana data-data tersebut diarsipkan pada satu *file* kemudian dicatat pada buku besar dan dimasukkan ke dalam komputer lalu disimpan sebagai arsip. Hal tersebut dirasakan akan menimbulkan permasalahan diantaranya resiko kehilangan data sangat besar, lambatnya proses pencarian data karena melibatkan banyak dokumen, rumitnya pemrosesan data, dan juga ada kemungkinan terjadinya kerangkapan data. Jadi apabila laporan penjualan dan perbaikan komputer dibutukan oleh pimpinan maka bagian administrasi merasa kesulitan dan informasi yang diberikanpun tidak bisa dipakai untuk penunjang keputusan bagi pimpinan.

Pada bagian perbaikan komputer di CV Computer Plus Palembang dalam melakukan *service* mempunyai kendala-kendala yaitu pada saat mau melakukan *service* komputer pihak teknisi harus melihat atau memeriksa terlebih dahulu persediaan barang sehingga membuang waktu karena apabila barang yang dibutuhkan teknisi tidak ada maka barang tersebut akan dipesan terlebih dahulu. Hal tersebut akan memperlambat kerja para teknisi dalam memperbaiki komputer. Begitu juga dengan pengolahan data *service* konputer sama seperti dengan pengolahan data penjualan yang terjadi sehari-hari.

Berdasarkan uraian permasalahan di atas, maka penulis mencoba membahas dan memberikan pemecahan masalah tersebut dengan judul "Sistem Informasi Penjualan dan Perbaikan Komputer Pada CV Computer Plus Palembang*"* dengan menggunakan *Microsoft Visual Basic 6.0.*

Adapun pokok permasalahan yang diangkat adalah "Bagaimana cara membuat Sistem Informasi Penjualan dan Perbaikan Komputer pada CV Computer Plus Palembang" dengan menggunakan pemrograman *Microsoft Visual Basic 6.0*. Dengan dibuatnya sistem informasi penjualan dan perbaikan komputer ini diharapkan agar dapat mempermudah dalam mengolah data transaksi penjualan dan perbaikan komputer yang terjadi setiap harinya. Begitu juga dalam pembuatan laporan penjualan dan perbaikan komputer serta mempermudah dalam pengecekan persediaan barang yang berada di gudang. Sehingga pengolahan data yang dilakukan akan lebih baik dan informasi yang dihasilkan akan lebih cepat, tepat dan akurat.

Tujuan Penelitian adalah: Membuat Sistem Informasi Penjualan dan Perbaikan Komputer yang berbasis komputer dengan menggunakan aplikasi pemrograman *Microsoft Visual Basic 6.0* pada CV Computer Plus Palembang. Sedangkan Manfaat Penelitian: Dapat membantu bagian penjualan dan perbaikan komputer pada CV Computer Plus Palembang dalam memproses pengolahan data transaksi penjualan dan perbaikan komputer yang terjadi setiap harinya.

Agar penelitian focus dan relevan, maka perlu dibatasi pengolahan data, yang meliputi data *supplier*, pelanggan, teknisi, pembelian, persediaan barang, penjualan, *service* komputer sehingga dapat menghasilkan informasi penjualan dan perbaikan komputer pada CV Computer Plus Palembang.

## 2. TINJAUAN PUSTAKA

Untuk memudahkan proses penelitian sistem informasi penjualan dan perbaikan, perlu diketahui sejumlah teori atau penelitian yang pernah dilakukan sebagai rujukan.

### 2.1 Sistem Informasi

Sutabri (2005:42) Sistem Informasi adalah suatu sistem di dalam suatu organisasi yang mempertemukan kebutuhan pengolahan transaksi harian yang mendukung fungsi operasi organisasi yang bersifat manajerial dengan kegiatan strategi dari suatu organisasi untuk dapat menyediakan kepada pihak luar tertentu dengan laporan-laporan yang diperlukan.

Sedangkan Abdillah (2004:134) berpendapat bahwa Secara sederhana, Sistem Informasi merupakan kumpulan komponen yang saling berhubungan untuk mengolah *input* (data) menjadi *output* (informasi) sehingga dapat memenuhi kebutuhan pemakai.

Dari pendapat-pendapat di atas dapat disimpulkan bahwa sistem informasi adalah suatu sistem yang terdiri dari kombinasi yang dibutuhkan oleh organisasi untuk mencapai tujuan (mengolah *input* menjadi *output*).

### 2.2 Penjualan

Suryana (2003:118) Penjualan adalah menyajikan barang agar konsumen menjadi tertarik dan melakukan pembelian. Penjulan dapat dilakukan dengan cara langsung mendatangi konsumen, menunggu kedatangan konsumen dan melayani konsumen.

### 2.3 Perbaikan Komputer

Anwar (2005:256) mengemukanan perbaikan adalah pembetulan, hal (hasil, perbuatan, usaha dan sebagainya) memperbaiki; di keadaan menjadi baik. Perihal berbaik kembali; Perubahan yang mengakibatkan penggunaan alat dapat lebih lama.

Sedangkan komputer (http://www.total.or.id/info.php?kk=computer) adalah hasil dari kemajuan teknologi elektronika dan informatika yang berfungsi sebagai alat bantu untuk menulis, menggambar, menyunting gambar atau foto, membuat animasi, mengoperasikan program analisis ilmiah, simulasi dan untuk kontrol peralatan.

Jadi Perbaikan Komputer adalah usaha dengan tujuan untuk membuat komputer dari keadaan yang tidak baik/belum baik/rusak menjadi baik kembali dalam arti mampu berfungsi kembali sebagaimana mestinya.

## 3. METODOLOGI PENELITIAN

### 3.1 Lokasi Penelitian

Lokasi penelitian dilakukan pada CV Computer Plus Palembang, beralamat di Jalan Madang Lr. Makmur V No. 47 Palembang. Pelaksanaan penelitian ini dimulai pada Bulan Mei dan berakhir pada Bulan Juli 2008.

### 3.2 Metode Pengumpulan Data

Untuk mendapatkan data dan informasi yang diperlukan, penulis menggunakan metode deskriptif, yaitu dengan cara mengumpulkan data dan informasi di CV Computer Plus Palembang. Penulis mengadakan penelitian dengan cara sebagai berikut: 1) Observasi, dengan melakukan pengamatan langsung pada CV Computer Plus, 2) Wawancara, dilakukan langsung dengan a) Pimpinan, b) Bagian Gudang, c) Teknisi, dan d) bagian Penjualan, serta 3) Studi Pustaka, dilakukan dengan membaca buku (Abdillah, 2004:137).

### 3.3 *Software* yang Dibutuhkan

Dalam membangun sistem informasi ini dibutuhkan sejumlah *software*, yaitu: 1) Microsoft Visual BASIC 6.0, 2) Microsoft Access XP, 3) Crystal Report, serta dibantu dengan 4) Microsoft Visio, dan 5) Microsoft Word.

### 3.4 Metode Pengembangan Sistem

Metode pengembangan sistem sangat dibutuhkan untuk mempermudah tahapan kegiatan yang akan peneliti lakukan. Sutedjo (2002:151) mengemukakan tahap-tahap atau siklus sistem yang digunakan dalam melakukan pengembangan sistem perangkat lunak adalah: 1) Perencanaan: Definisi masalah pada tahap ini dilakukan kajian untuk memperoleh gambaran permasalahan dan ruang lingkupnya yang terdapat pada sistem yang akan dikembangkan. Setelah permasalahan ditemukan diharapkan solusi untuk mengatasi permasalahan tersebut dapat diperoleh, 2) Analisa: Pada tahap ini permasalahan dianalisis secara lebih mendalam dengan mempelajari sistem yang sedang berjalan untuk membantu sistem yang baru dan menentukan kebutuhan informasi pemakai. Dari sistem ini akan dapat disimpulkan bahwa sistem yang baru layak atau tidak untuk dikembangkan, 3) Perancangan: Setelah mengetahui sistem yang lama dan mengetahui kriteria-kriteria sistem yang dibangun kemudian dibuatlah desain masukan, desain proses dan desain keluaran serta desain *database*-nya, 4) Penerapan: Hasil perancangan dari tahap sebelumnya diimplementasikan dibagian ini untuk menerapkan prosedur dalam teknologi komputer digunakan aplikasi pemrograman, sedangkan untuk proses yang berada diluar sistem komputer disusunlah suatu aturan agar setiap orang yang terlibat dapat mengikuti prosedur yang telah ditentukan, dan 6) Evaluasi: Setelah sistem dibuat maka dilakukan pengujian terhadap sistem tersebut. Pengujian dilakukan untuk mengetahui proses pengolahan dan menjadi informasi dan mendistribusikannya.

### 3.5 Diagram Konteks Sistem Infromasi Penjualan dan Perbaikan Komputer

*Entity* yang terlibat pada sistem informasi ini adalah: 1) Pelanggan, 2) Teknisi, 3) Pemasok, 4) Gudang, dan 5) Pimpinan. Secara umum Diagram Konteks Sistem Informasi ini dapat dilihat pada Gambar 1.

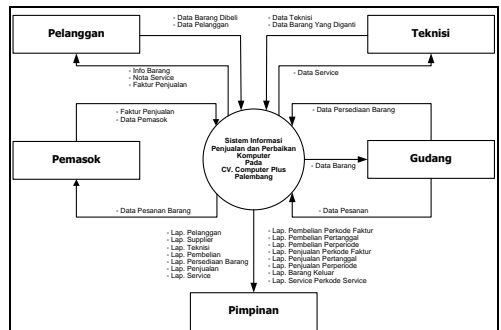

**Gambar 1. Diagram Konteks SI Penjualan dan Perbaikan**

## 4. PEMBAHASAN

Setelah semua tahapan pengembangan sistem dilakukan, maka didapatlah suatu perangkat lunak sistem informasi penjualan dan perbaikan komputer, yang terdiri atas sejumlah *form* dan *report*.

### 4.1 Menu Utama

Menu utama adalah form yang pertama kali tampil yang berfungsi untuk mengorganisasikan submenu-submenu dibawahnya yang saling berhubungan. Menu Utama terdiri atas 4 (empat) Menu. Tampilan menu utama dapat dilihat pada gambar berikut ini:

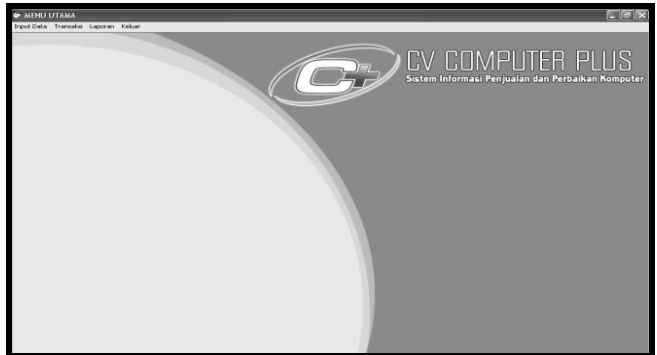

**Gambar 1. Tampilan Menu Utama**

### 4.2 Menu Data Pelanggan

Digunakan untuk mencatat dan melihat data-data pelanggan. Dipanggil dengan meng-klik input data pelanggan dari menu utama. Untuk mengisi data pelanggan klik tombol *new* (jadi tombol *save*). Kode pelanggan muncul otomatis.

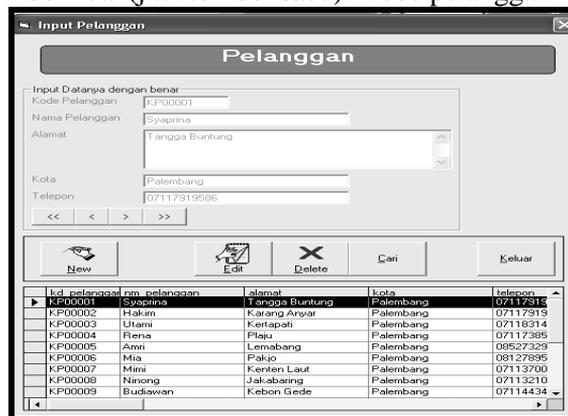

**Gambar 2. Tampilan Input Data Pelanggan**

### 4.3 Menu Data *Supplier*

Digunakan untuk mencatat dan melihat data-data *supplier*. Untuk melakukan pengisian data *supplier* dengan cara mengklik pada input data *supplier* dari menu utama.

**Gambar 3. Tampilan Input Data *Supplier***

### 4.4 Menu Data Teknisi

Digunakan untuk mencatat dan melihat data-data teknisi. Dipanggil dengan meng-klik input data teknisi dari menu utama. Untuk mengisi data teknisi, klik tombol *new* (menjadi tombol *save*). Kode teknisi akan muncul secara otomatis. Lihat gambar berikut.

**Gambar 4. Tampilan Input Data Teknisi**

### 4.5 Menu Data Persediaan Barang

Digunakan untuk mencatat dan melihat data-data persediaan barang. Untuk melakukan pilihan ini adalah dengan cara mengklik pada input data persediaan barang dari menu utama, kemudian akan tampil sub menu didalamya.

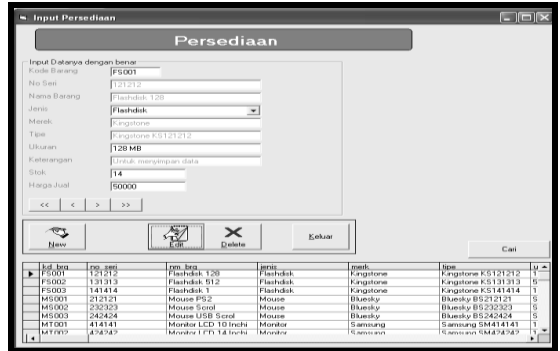
**Gambar 5. Tampilan Input Data Persediaan**

### 4.6 Menu Data Pembelian

Digunakan untuk mencatat dan melihat data-data pembelian dari supplier. Untuk melakukan pilihan ini adalah dengan cara mengklik pada input data transaksi pembelian dari menu utama, kemudian akan tampil sub menu didalamnya. Pada menu pembelian terdapat sejumlah *fields* yang harus diisi. Tampilan form data pembelian seperti gambar berikut:

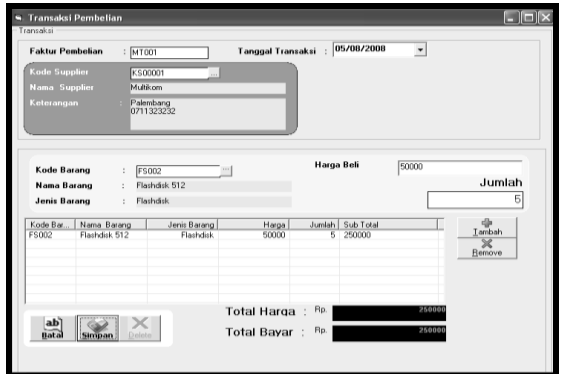
**Gambar 6. Tampilan Input Data Pembelian**

### 4.7 Menu Data Penjualan

Digunakan untuk mencatat dan melihat data-data penjualan. Untuk melakukan pilihan ini adalah dengan cara mengklik pada *input* data transaksi penjualan dari menu utama, kemudian akan tampil sub menu didalamnya. Pada menu penjualan terdapat *field-field* yang harus diisi. Ketika mengklik tobol *New* maka kode fakturnya akan tampil secara otomatis.

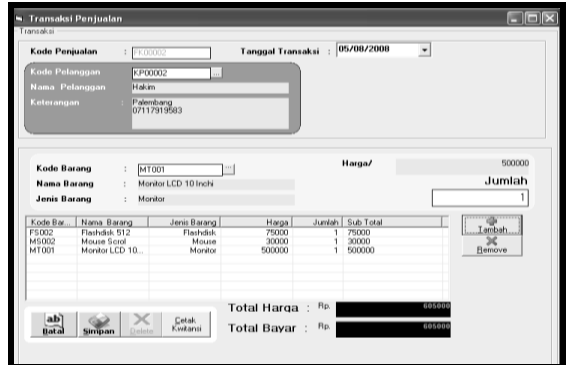

**Gambar 7. Tampilan Input Data Penjualan**

## 4.8 Menu Data Perbaikan (Terima Barang)

Maksudnya pada saat pelanggan mau memperbaiki komputer atau sejenisnya maka pelanggan terlebih dahulu didata dan untuk tanda terimanya pelanggan akan mendapat kwitasi *service* dan CV Computer Plus menerima barang yang akan di*service* tadi. Tampilan form terima barang sebagai berikut:

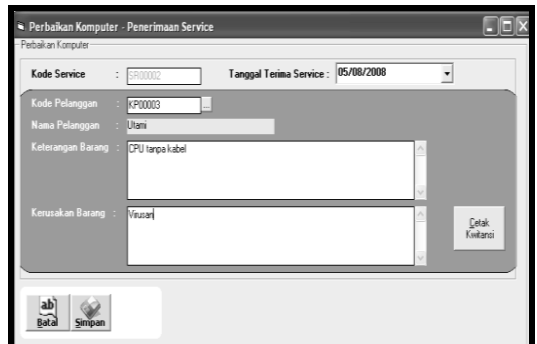

**Gambar 8.** Tampilan Input Terima Barang

## 4.9 Menu Data Perbaikan (Ambil Barang)

Maksudnya pada saat pelanggan ingin mengambil barang yang *diservice* maka pelanggan memberikan kwitansi kepada bagian administrasi dan pelanggan akan mendapatkan kwitasi pembayaran *service* sekalian barang yang telah selesai. Tampilan dapat dilihat sebagai berikut

**Gambar 9. Tampilan Input Ambil Barang**

## 4.10 Laporan Penjualan

Laporan ini akan menampilkan hasil pengolahan data penjualan. Untuk mengetahui laporan penjualan dapat dilakukan dengan klik submenu 'laporan' pada menu utama lalu klik "Laporan Penjualan".

**Gambar 10. Laporan Penjualan**

## 4.11 Laporan Perbaikan *(Service)*

Laporan ini akan menampilkan hasil pengolahan data perbaikan. Untuk mengetahui laporan perbaikan dapat dilakukan dengan klik submenu 'laporan' pada menu utama lalu klik "Laporan Perbaikan". Sebagai contoh lihat gambar berikut:

**Gambar 11. Laporan Perbaikan** *(Service)*

### 4.12 Analisis Pembahasan

Semua kemudahan disiapkan dengan menggunakan suatu Menu Utama yang dapat memanggil sub program sehingga memudahkan petugas untuk melakukan operasi atas data Penjualan dan Perbaikan Komputer Pada CV Computer Plus Palembang.

Setiap *form* diberikan fasilitas berupa *AutoNumber* sehingga *primary key* secara otomatis akan diberikan secara Increment, hal ini dapat mempercepat aktivitas input data oleh petugas, memudahkan dalam mengecek persediaan barang pada gudang.

*Report* juga dibuat dengan sekali klik petugas dapat melihat dan mencetak report dengan tampilan yang menarik serta dilengkapi beragam konfigurasi laporannya.

### 5. SIMPULAN

Berdasarkan hasil penelitian, analisis, serta pembahasan sistem informasi penjualan dan perbaikan komputer pada CV Computer PLUS, dapat ditarik sejumlah kesimpulan, sebagai berikut:
1) Aplikasi yang dihasilkan adalah Sistem Informasi Penjualan dan Perbaikan Komputer Pada CV Computer Plus Palembang.
2) Memberikan kemudahan pada bagian administrasi dalam mengolah data penjualan dan perbaikan komputer pada CV Computer Plus Palembang.

Serta dapat memudahkan dalam mengecek persediaan barang pada gudang.
3) Aplikasi yang dihasilkan dapat mendukung dan mempercepat dalam pengolahan data-data penjualan dan perbaikan komputer dan dapat meningkatkan efektifitas kinerja pada CV Computer Plus Palembang.
4) Untuk mengoperasikan sistem yang dibangun diperlukan seseorang yang memiliki kemampuan minimal dapat mengoperasikan aplikasi komputer visual.

# DAFTAR RUJUKAN